\documentclass[aps,epsfig,floatfix,twocolumn,nofootinbib,showpacs]{revtex4}
\usepackage{graphicx}
\usepackage{latexsym}

\begin{document}
\draft

\title{ Space-Charge-Limited Current Fluctuations in Organic
Semiconductors
 }

\author{ A. Carbone$^1$,  B. K. Kotowska$^{1,2}$, D.  Kotowski$^{1,2}$}

\affiliation{ $^1$ Physics Department and National Institute of
Matter Physics (INFM), Politecnico
di Torino, C.so Duca degli Abruzzi 24, 10129 Torino, Italy \\
$^2$ Department of Physics of Electronic Phenomena, Gdansk
University of Technology, Narutowicza 11/12, 80-952 Gdansk,
Poland  }



\begin{abstract}
 Low-frequency current fluctuations are investigated over a bias range covering {\em ohmic},
{\em trap-filling} and {\em space-charge-limited  current} regimes
in polycrystalline polyacenes.  The relative current noise power
spectral density ${\cal S}(f)$ is constant in the {\em ohmic }
region, steeply increases at the {\em trap-filling transition}
region and decreases in the {\em space-charge-limited-current}
region.
   The {\em noise peak} at the {\em  trap-filling
transition} is accounted for within a {\em continuum percolation
model}. As the quasi-Fermi level crosses the trap level, intricate
insulating paths nucleate within the ohmic matrix, determining the
onset of non-equilibrium conditions at the interface between the
insulating and conducting phase. The {\em noise peak} is written
in terms of the free and trapped charge carrier densities.

\end{abstract}

\pacs{73.50.Ph, 72.70.+m, 72.80.Le, 72.80.Ng }

\maketitle Polycrystalline small-weight organic materials, as
polyacenes, belong to the class of strongly disordered conductors.
The charge carrier transport mainly occurs by variable-range
 hopping among a system of localized electronic states
 \cite{Pope,Parris,Fishchuk}, critically
  depends on the injection from the
  metal electrode   \cite{Shen,Hegmann,Reichert,Silveira} and is
  affected by the presence of deep and shallow traps at the metal-organic interface  and in the
 bulk \cite{Koch,Deboer,Arkhipov}. Compared to inorganic materials,
 the identification and characterization of defects in organic
 semiconductors is a more recent issue
 \cite{Yang,Knipp,Northrup,Lang,Muller,Kang}. Defects are often related to chemical impurity, e.g. anthracene
as impurity in tetracene/pentacene. A general model, based on
density functional calculations of gap states generated  by
hydrogen or oxygen impurities in a $C-H$ unit valid for small and
long chain molecular materials, has been proposed \cite{Northrup}.
A metastable defect generation phenomenon driven by bias has been
observed by using space-charge-limited current spectroscopy
\cite{Lang}.
    Long-lived deep traps, located in the grains and
evolving with voltage, have been imaged by electric force
microscopy \cite{Muller}. Shallow traps originated by the sliding
of pentacene molecules have been recently observed in \cite{Kang}.
\par Noise studies
have been so far addressed to devices performances
\cite{SampietroNecliudovMartin} rather than to carrier dynamics in
organic semiconductors. However, current fluctuations can provide
information
 about nonuniform charge distributions and
 meandering current flow paths arising in the presence of disorder
 \cite{Kogan,Shklovskii,Rammal,Bardhan,Carbone,Elteto}.
 The
 emergence of disorder determines non-equilibrium conditions at the
interface between different phases in systems exhibiting nonlinear
response to external fields and threshold behavior to the onset of
a steady-state. Such systems, e.g. flux lines in disordered
superconductors, charge density waves pinned by impurities, phase
separation in manganites, charge tunnel in metal dot arrays
\cite{Elteto}, share the feature that the transition from a weakly
disordered state - characterized by steady fluctuations - to a
strongly disordered state - characterized by critical fluctuations
- is driven by a bias. In this Letter, we report the first study
of fluctuations over three transport regimes - {\em ohmic}, {\em
trap-filling}, {\em space-charge-limited} - in polyacenes. We
observe that the relative
 power spectral density ${\cal S}{(f)}=S_I(f)/I^2$ \cite{notea}
 is consistent  with {\em steady-state fluctuations} in ohmic regime. At the trap-filling
transition (TFT) between  ohmic and insulating regime, we measure
a rapid increase of ${\cal S}{(f)}$. Beyond the {\em threshold
voltage} $V_T$, at the onset of the space-charge-limited regime,
${\cal S}{(f)}$ decreases, as expected for steady SCLC
fluctuations. The strong increase of ${\cal S}{(f)}$ is discussed
within a percolative fluctuations model and is related to the
 conductor-insulator interface instabilities when the
insulating domains increase at the expenses of the conductive
ones. The ${\cal S}{(f)}$ peak is written in terms of the
trapped-to-free charge carrier ratio, directly related to the
insulating-conductive phase imbalance. Finally, a mechanism of
trap formation due to bias-stress \cite{Northrup,Lang,Muller}
accounts for the progressive divergence of ${\cal S}{(f)}$
preceding the breakdown.

\par
\begin{figure}

\includegraphics[width=7cm,height=5cm,angle=0]{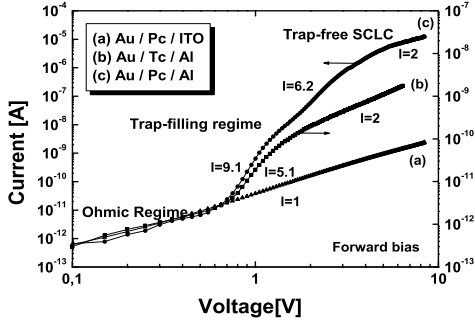}
\caption{\label{IV} Current-voltage characteristics for  (a)
Au/Pc/ITO with $L=0.85 \mu m$, (b) Au/Tc/Al with $L=0.65 \mu m$,
(c) Au/Pc/Al with $L=0.85 \mu m$. The geometry is planar with
sandwich configuration of the electrodes. Curve (a) is linear over
all the investigated voltage range ($l=1$) and is given in
arbitrary units. Curve (b) exhibits the typical SCLC behavior
(ohmic regime$\Rightarrow$trap-filling transition$\Rightarrow$SCLC
regime). Curve (c) exhibits a more complicated behavior very
likely related to the deep traps distributed around two different
energy levels }
\end{figure}

 Pentacene
$C_{22}H_{14}$ and tetracene $C_{18}H_{12}$ purified by
sublimation have been evaporated on glass at $10^{-5}$Pa and room
temperature. Sandwich structures with Au, Al and ITO electrodes,
with area $A=0.1 cm^2$, distant $L=0.40\div 1.00 \mu m$ with
$\delta L=0.05 \mu m$, are investigated. This large set guarantees
a reliable statistics over the variations of chemical purity
degree and structural homogeneity. Current-voltage $I-V$ curves
are shown in Figure~\ref{IV} for: (a) Au/Pc/ITO, (b) Au/Tc/Al, (c)
Au/Pc/Al. Curve (a) is linear over all the investigated range.
Curves (b) and (c) show the typical shape of space-charge-limited
current in materials with deep traps. The slope $l=1$ refers to
the {\em ohmic regime}, described by $ J_{\Omega}=q \mu n V/L $.
The regions with steep slope, {\em trap filling regime},
correspond to the rapid change undergone by the current as the
Fermi level $E_F$ moves through a trap level $E_t$. The slope
$l=2$, refers to the {\em trap-free space-charge-limited-current}
regime, obeying the Mott-Gurney law $ J_{SCLC}=9
\epsilon\epsilon_0\mu \Theta {V^2}/{8 L^3}$ \cite{Pope}.

\begin{figure}
\includegraphics[width=7cm,height=5cm,angle=0]{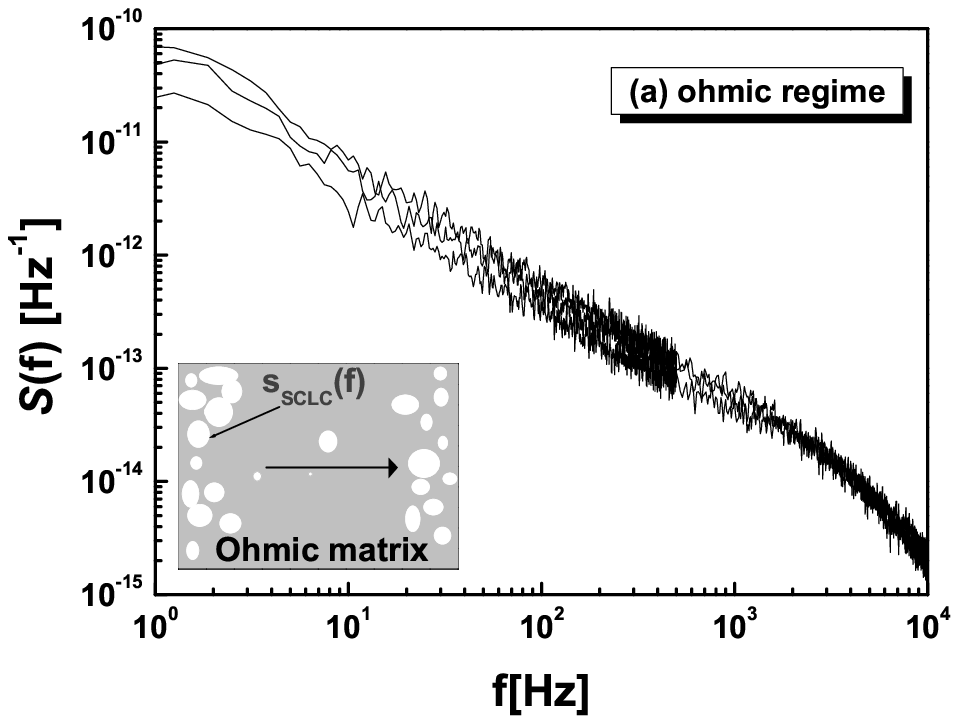}
\includegraphics[width=7cm,height=5cm,angle=0]{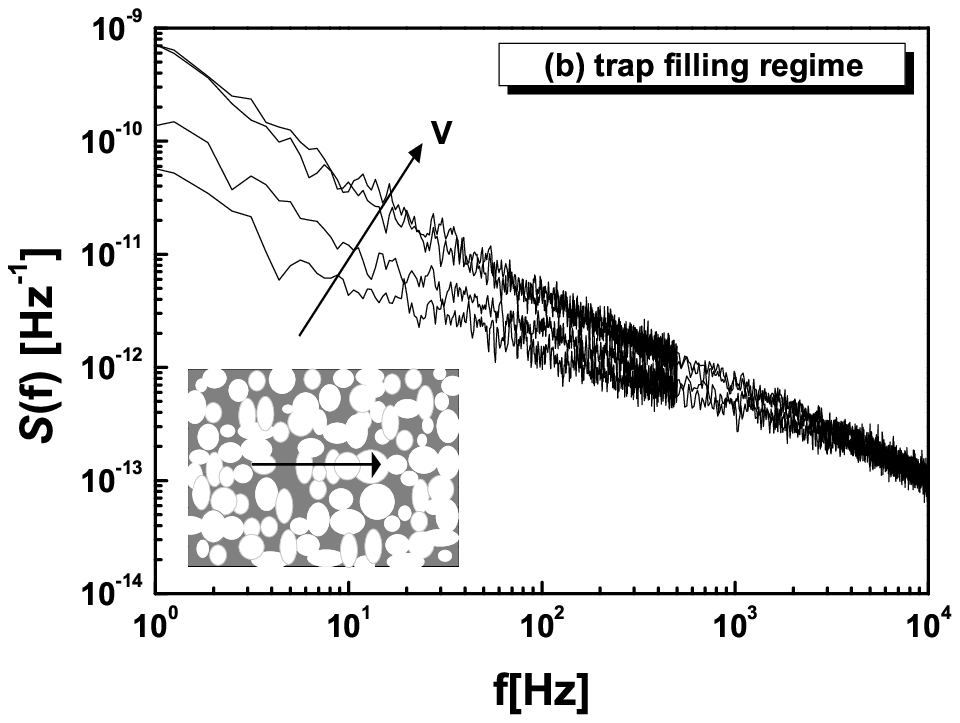}
\includegraphics[width=7cm,height=5cm,angle=0]{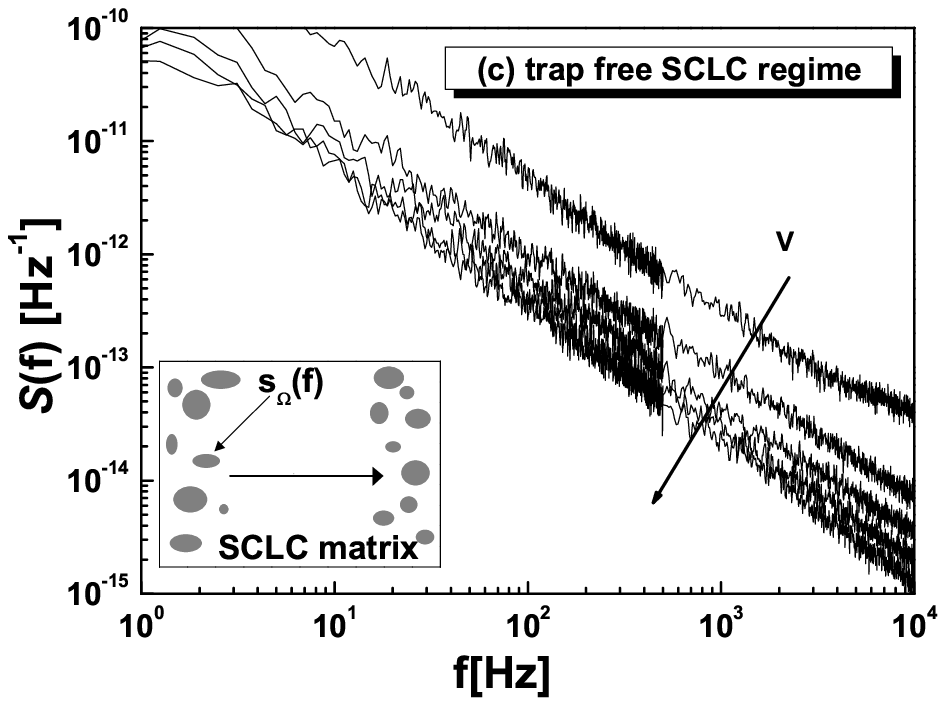}
\caption{\label{Noise} Relative current noise power spectral
density ${\cal S}{(f)}$ for the Au/Tc/Al sample. (a) {\em Ohmic
regime}: ${\cal S}{(f)}$ does not vary with $V$ in ohmic condition
($0.3\div 0.8$ V). (b) {\em Trap-filling regime}: ${\cal S}{(f)}$
sharply increases with V during the trap filling transition ($0.8
\div 2$ V). (c) {\em Space-charge-limited current regime}: ${\cal
S}{(f)}$ decreases approximately as $1/V$ ( $>2$ V).   The
two-phases medium is shown in the insets. The horizontal arrows
represent the current direction. The white areas represent filled
traps, i.e. insulating sites characterized by $s_{SCLC}(f)$ noise.
The dark areas represent empty traps, i.e. conductive sites
characterized by $s_{\Omega}(f)$ noise. (a) The quasi-Fermi level
$E_F \ll E_t$, almost all deep traps are empty. Transport is
ohmic. (b) The quasi-Fermi level is moving through the trap level,
$E_F \sim E_t$, filling the traps. Tortuous insulating patterns
are generated inside the conductive matrix, leading to
non-equilibrium condition at the ohmic-insulating interface and
excess fluctuations. (c) The quasi-Fermi level $E_F \gg E_t$, the
traps are mostly filled. The system is characterized by
steady-state space-charge-limited current fluctuations decreasing
with $V$. Darker areas represent residual conductive sites,
shallower tails of gaussian distributed traps. }
\end{figure}

\begin{figure}
\includegraphics[width=7cm,height=5cm,angle=0]{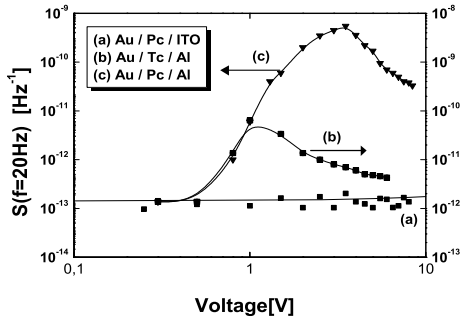}
\caption{\label{Noiseall} Log-log plot of the voltage dependence
of the relative fluctuation power spectral density at frequency
$f=20Hz$ for (a) Au/Pc/ITO, (b) Au/Tc/Al, (c) Au/Pc/Al. Data (a)
are given in arbitrary units, mostly for reference purpose,  and
correspond to ohmic behavior over all the voltage range. The I-V
characteristics for the same samples are plotted in Fig.~\ref{IV}}

\end{figure}

\par
 Relative current noise power spectra ${\cal S}{(f)}$ are shown in Figure~\ref{Noise}
  for the $Au/Tc/Al$ sample respectively in (a) {\em ohmic}, (b)
{\em trap-filling transition} and (c) {\em space-charge-limited
current}
 regime, at room temperature, in the dark. The frequency
dependence is $f^{-\gamma}$ with ${\gamma}\approx 1 $. The signal
is acquired, Fourier-transformed and 50 times averaged over the
ranges $1\div 500Hz$ and $500\div10^4Hz$.
 The continuity  of the low and high frequency branches
 ensures the noise stationarity.
 Roll-off and saturation of ${\cal S}{(f)}$ at $f>1kHz$
are  respectively due to unavoidable capacitive coupling
 and  circuitry background noise with the low currents ($I < 10^{-10}A$)  and high
 resistances ($R> 10^{4}M\Omega$)  into play. Therefore, our
 discussion is limited to the frequency range where the
 $f^{-\gamma}$ component dominates over the other noise sources. In
Fig.~\ref{Noise}~(a), ${\cal S}{(f)}$ does not change with
voltage, as expected for uncorrelated resistance fluctuations in
nearly ideal ohmic conditions \cite{Kogan}. Fig.~\ref{Noise}(b)
refers to the {\em trap-filling regime}: ${\cal S}{(f)}$ sharply
increases with $V$. Fig.~\ref{Noise}~(c) refers to the trap-free
SCLC region: ${\cal S}{(f)}$ decreases approximately as $1/V$
according to a noise suppression mechanism analogous to that
observed in vacuum tubes and inorganic solid-state diodes
operating under space-charge-limited conditions
\cite{Kleinpenning}.
 These results are summarized in Figure~\ref{Noiseall},
where ${\cal S}{(f=20Hz)}$ is plotted over the entire range of
voltage for the samples of Fig.~\ref{IV}. The striking feature is
the peak exhibited by the relative noise intensity ${\cal S}(f)$
in the samples undergoing the trap filling transition. The
presence of the peak is the unequivocal signature of
nonequilibrium and strong correlation effects causing
multiplicative mechanisms of noise generation.
\par Here we provide an interpretation of the noise results based on a percolation model.
 In the ohmic regime,
 the conductive component almost exclusively consists of
 thermally excited charge carriers. The deep traps are mostly empty ($\Omega$ {\em
 phase}). In the space-charge-limited current regime, the transport is dominated by the injected
holes controlled by space-charge. The deep traps are almost
completely filled (SCLC {\em phase}). In the intermediate voltage
region, {\em trap-filling transition}, the system can be viewed as
a two-components continuum percolative medium \cite{Stanley}
characterized by the {\em competition} between the conductive
($\Omega$) and the insulating (SCLC) phase driven by voltage. The
{\em ohmic phase} becomes populated by insulating sites as the
voltage increases. The current paths are extremely intricate owing
to the inhomogeneous distribution of trapping centers, whose
occupancy randomly evolves as the Fermi level moves through the
trap level. The system is in a strongly disordered critical state,
due to the nucleation of insulating patterns inside the conductive
medium. By further voltage increase beyond the threshold $V_T$,
steady state SCLC fluctuations decreasing with $V$ are observed.
The increase of fluctuations, at the TFT transition, is related to
the greatly disordered distribution of local fields, compared to
the more ordered distribution in ohmic and SCLC regimes.  The
competition between repulsive and attractive Coulomb interaction
undergone by the carriers moving through oppositely charged sites
determines strong correlation effects among the elementary hop
instances. The fluctuations of such a system cannot be described
as a simple sum of the noise terms related to the $\Omega$  and
SCLC regions. Let $s_\Omega(f)$ and $s_{SCLC}(f)$ indicate the
noise sources characterizing respectively the conductive and the
insulating elementary sites. An estimate of the noise peak  when
the percolative regime is approached from the conductive side can
be obtained
 using the relationship
  \cite{Rammal}:
\begin{equation}
\label{percolativenoise} {\cal
S}(f)={s}_{\Omega}(f)\frac{\sum_\alpha i^4_\alpha}{(\sum_\alpha
i^2_\alpha)^2} \hspace{10pt},
\end{equation}
where ${s}_{\Omega}(f)$ and $i_\alpha$ indicate respectively the
spectral density and the current of each element of the conductive
network. The frequency dependence of ${\cal S}(f)$ is contained in
the first factor of Eq.(\ref{percolativenoise}). According to the
noise models for variable range hopping \cite{Kogan,Shklovskii},
${s}_{\Omega}(f)$  can be expected to be $f ^{-\gamma}$ sloped
with $\gamma\approx 1$. In our system, the hop instances are
kicked off respectively by the thermally activated
  detrapping  ($\Omega$ regime) or by the injection (SCLC regime) of a charge carrier.
  The fluctuation amplitude is determined by the last factor of
Eq.(\ref{percolativenoise}), that is related to the conductive
volume fraction $\phi$ by:
\begin{equation}
\label{percolativeS}
 {\cal S} \propto \Delta\phi^{-k} \hspace{10pt},
\end{equation}
where $k$ is a critical exponent, whose value depends  on the
structure, composition and conduction mechanism  \cite{Note}. The
conductive fraction $\phi$ depends on $V$. Moreover, ought to a
possible mechanism of deep trap formation by bias/thermal stress
\cite{Knipp,Lang,Muller,Northrup},  the total density of trap
$N_t$ increases. Thus the definition of an universal value of
 $k$ remains elusive. In our samples, $k$ ranges from $1.1$ to
 $1.8$ under strict-sense stationary noise conditions.
 The change of conductive fraction due to the filling of deep traps can be written as $\Delta \phi \propto ({n-n_t})/{N_v}$, where $n$
and $n_t$ are respectively the free and trapped charge carrier
density, ${N_v}$ is the total density of states, coinciding with
the molecular density for narrow band materials. Since the
relative noise intensity for ohmic conductors varies as $1/n$, it
is convenient to write $\Delta \phi$ as:
\begin{equation}
\label{pcnoiseNt} \Delta \phi \propto \frac{n}{N_v} \left( 1-
\frac{n_t}{n}\right)\hspace{10pt} .
\end{equation}
By substitution of  Eq.~(\ref{pcnoiseNt}) into
Eq.~(\ref{percolativeS}), it follows that the noise in excess with
respect to the ohmic level is determined by the quantity
$(1-{n_t}/{n})$. It is related to the imbalance between free and
trapped carriers and, ultimately, to the departure from the
quasiequilibrium ohmic condition. Assuming for simplicity a
discrete trap level, it is $n=N_v \exp[-(E_v-E_F)/kT]$ and
$n_t={N_t}/{\{1+g^{-1}\exp[-(E_F-E_t)/kT]\}} \simeq {2
N_t}\exp[(E_F-E_t)/kT]$,   $g$ being the trap degeneracy factor of
and the other quantities introduced above. The ratio ${n_t}/{n}$
is written:
\begin{equation}
\label{pcnoiseNt2}  \frac{n_t}{n}= \frac{2
N_t\exp{[-(E_t-E_v)}/{kT}]}{ N_v} \hspace{10pt}.
\end{equation}
\par

Moreover, the Eq.~(\ref{pcnoiseNt2}) relates the steep increase of
${\cal S}(f)$  to that of the current upon trap filling
\cite{Pope,Deboer}, providing an independent validation of the
proposed noise picture being the conductivity variation
proportional to $ \Delta\phi ^{-\rho}$ within the percolative
model. Since $N_v=4\cdot10^{21}cm^{-3}$, $E_t$ and $N_t$ typically
range between $0.3\div
 0.6eV$
and  $10^{15}\div 10^{18}cm^{-3}$,
  the  Eq.~(\ref{pcnoiseNt2})
confirms that  a very small density of deep traps
 may critically affect the fluctuations. The
percolation threshold $\phi_c$ and the onset of breakdown are
finally discussed using
Eqs.~(\ref{percolativeS}-\ref{pcnoiseNt2}). Non-stationarity
 and  divergence of noise at the
trap-filling transition are the precursors of electrical
breakdown. Using the Eq.~(\ref{pcnoiseNt2}), the
 percolative threshold
$\phi_c$ is reached when $\Delta \phi \Rightarrow 0 $ i.e. ${ 2
N_t\exp{[-(E_t-E_v)}/{kT}]}\Rightarrow {N_v}$. The noise
divergence, observed after several bias or thermal cycles, might
be  caused by the increase of $N_t$ due to the deep trap formation
mechanism  suggested in \cite{Lang,Muller,Northrup}.
\par
  In conclusion, we have observed: (i)  {\em steady-state fluctuations} at
low-voltage (ohmic regime); (ii) {\em critical fluctuations} at
intermediate voltage (TFT transition); (iii) {\em steady-state
fluctuations} at high voltage (space-charge-limited-current
regime).  The $f^{-\gamma }$ shape indicates that the fluctuations
result from hops driven by trapping-detrapping processes with a
broad range of characteristic times $\tau$, in agreement with
\cite{Muller}. The noise peak, at the TFT transition, has been
ascribed to the strong nonequilibrium provoked by the insulating
phase nucleating within the conductive one. The noise peak has
been estimated within a percolation model of fluctuations:
Eqs.~(\ref{pcnoiseNt}-\ref{pcnoiseNt2}).  These equations have
been used to relate the onset of breakdown to the percolation
threshold $\phi_c$. Finally, it is worthy to remark that: (1)
    the stochastic processes by which systems with
distributed thresholds undergo a transition
 driven by an external bias \cite{Elteto}; (2) the time-averaged processes
underlying space-charge-limited transport -  as for example the
the
 Goodman and Rose law predicted in 1971
  and the {\em seeming simple} 2D  planar emission \cite{Lebowitz} -  still represent open fundamental issues. In
this Letter, we have shown that a deep-insight can be achieved
across these two topics studying fluctuations at the
voltage-driven transition from ohmic to insulating phase in
space-charge-limited conditions.

\par We acknowledge  J. Godlewski,  G. Kaniadakis, A. N. Korotkov, P.
Mazzetti for reading the manuscript and
 MIUR and  MAE, contracts
PRIN2003029008 and 22-FI-2004-2006, for support.


\begin{references}
\bibitem{notea} $S_I(f)$ refers to the square modulus of the Fourier transform of the instantaneous fluctuations $\delta I(t)$ of the
current $I(t)$ around its average value $I$.
\end{references}
\end{document}